\documentclass[aps,prb,twocolumn,superscriptaddress,amsmath,showpacs]{revtex4}
\usepackage{graphicx}
\usepackage{bm}
\usepackage{times}
\begin{document}
\title{Fundamental mechanism underlying subwavelength optics of 
metamaterials: Charge oscillation-induced light emission and interference}
\author{X.~R.~Huang}
\email[]{xhuang@bnl.gov} 
\affiliation{National Synchrotron Light Source II, Brookhaven National 
Laboratory, Upton, New York 11973, USA}
\author{R.~W.~Peng}
\affiliation{National Laboratory of Solid State Microstructures, 
Nanjing University, Nanjing 210093, China}
\author{Mu Wang}
\affiliation{National Laboratory of Solid State Microstructures, 
Nanjing University, Nanjing 210093, China}

\begin{abstract}
Interactions between light and conducting nanostructures can 
result in a variety of novel and fascinating phenomena. These 
properties may have wide applications, but their underlying mechanisms 
have not been completely understood. From calculations of surface charge 
density waves on conducting gratings and by comparing them with 
classical surface plasmons, we revealed a general yet concrete picture 
about coupling of light to free electron oscillation on structured 
conducting surfaces that can lead to oscillating subwavelength charge patterns 
(i.e., spoof surface plasmons 
but without the dispersion property of classical surface plasmons). 
New wavelets emitted from these light sources 
then destructively interfere to form evanescent waves.
This principle, usually combined with 
other mechanisms (e.g. resonance), is mainly a 
geometrical effect that can be universally 
involved in light scattering from all periodic and nonperiodic structures 
containing free electrons, including perfect conductors. 
The spoof surface plasmon picture may provide clear guidelines for developing 
metamaterial-based nano-optical devices.
\end{abstract}
\maketitle

\section{Introduction}
\label{SecI}

The various novel and unusual optical properties of conducting nanostructures, 
such as anomalous diffraction from metallic gratings, enhanced light 
transmission through subwavelength slits or holes, 
light polarizing through wire grid polarizers,
surface-enhanced Raman scattering, negative refraction imaging, etc, 
have attracted tremendous attention in recent 
years.\cite{r1, r2, r3, WGP, ZhangX} 
To date the coupling of light with surface plasmons (SPs) have been widely 
adopted to explain these anomalous phenomena.
However, the SP picture elaborated in numerous case studies in the literature 
actually corresponds to a very general concept about coupling of 
electromagnetic (EM) waves to free electron oscillation on conducting 
surfaces that can generate evanescent EM 
wave modes. This big picture is correct without doubt, but it is too 
general for one to obtain a clear and straightforward 
understanding of the essential underlying mechanism.
Due to this uncertainty, the SP-like wave modes have been usually assumed to 
be similar to the classical SPs (CSPs) on planar metal surfaces,\cite{CSP} 
but this assumption is obviously 
challenged by the fact that (nearly) perfectly conducting structures that 
do not support CSPs still have 
\emph{similar but stronger\/} anomalous light scattering properties.\cite{Nanowire2} 
Conductors with positive permittivity do not support CSPs either,
but they can also exhibit light transmission anomalies.\cite{Lezec1,PRA,Sarrazin} 
(Extraordinary transmission through gratings can even occur for acoustic 
waves,\cite{Ming} which is completely irrelevant to SPs.) 
Because of these contradictions, the origin of 
anomalous light scattering from metallic nanostructures is still 
being argued (e.g. Refs. \onlinecite{Lezec2}-\onlinecite{NGarcia}).

Using modern computing techniques one may numerically solve Maxwell's 
equations for various complicated structures, but such computations have been 
largely focused on the EM fields. 
Surprisingly, the detailed mechanisms of free electron oscillation have been 
almost completely ignored in the literature although they are known to play 
the essential role in the SP picture. 
Very recently, we have briefly reported our computations of 
surface charge density waves (SCDWs) and the role they play 
in the process of enhanced light transmission through 
slit and hole arrays.\cite{PRA} 
In this paper, we give a detailed and comprehensive 
illustration of the basic mechanism about
light emission and interference from incident-wave-driven free electron 
oscillations, demonstrate that it is involved in light scattering from all 
periodic and nonperiodic 
conducting structures (including perfect conductors), 
and thus establish a simple, concrete and universal spoof SP picture. 
This picture 
may provide solid guidelines for designing nano-optical devices by directing 
people to concentrate on the geometrical parameters of conducting 
nanostructures so as to control the locations, strengths, and interference of 
the charge oscillation-induced light sources.

\section{charge oscillation-induced light emission and interference}
\label{SecII}

To illustrate the main picture, we start from the well-known 
principle of Thomson scattering of x-rays by electrons,\cite{XrayBook} in 
which the incident x rays 
(EM waves with wavelengths $\sim 0.1$ nm) 
force the electrons in atoms (not necessarily free electrons) 
to oscillate with the same frequency. 
According to the fact that \emph{accelerating charges radiate},
the oscillating electrons then emit new wavelets, 
which form the scattered waves. 
In principle, this effect also exists in the long wavelength range 
(say $\lambda>0.1$ $\mu$m), where electrons still oscillate with the incident wave 
(giving rise to oscillating polarization of the atoms). 
However, since now $\lambda$ is much larger than the atoms ($\sim 0.1$ nm), 
the net charge density averaged on the wavelength scale is zero in the bulk. 
Net polarization-induced charges do exist on surfaces (or interfaces), but for 
non-conducting materials where electrons are bound to atoms and cannot move 
freely, the formation of net oscillating charges 
is very small even on rough surfaces. 

\begin{figure} 
\includegraphics[scale=1.0,angle=0]{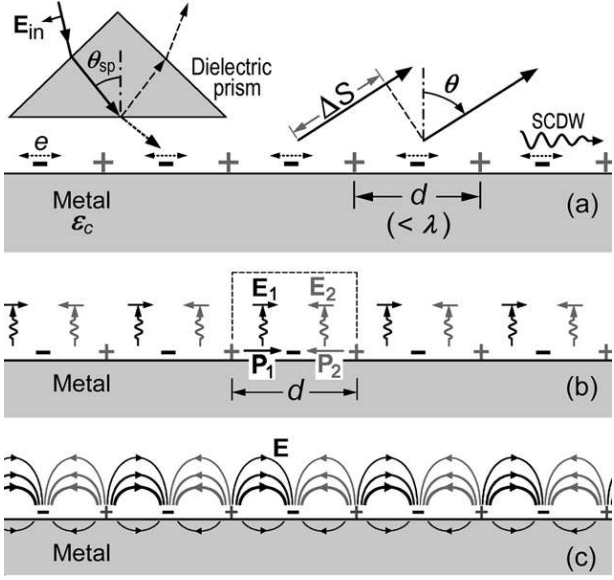}
\caption{CSP on a planar metal surface. 
(a) Excitation of the CSP by Otto geometry. 
The SCDW is the result of electron oscillations 
(indicated by the dashed arrows) 
while the positive charges are fixed. 
(b) Symmetric sub-wavelets (${\mathbf E}_1$ and ${\mathbf E}_2$) from a period 
of the charge wave (outlined by the dashed lines) along $\theta = 0$. 
${\mathbf P}_1$ and ${\mathbf P}_2$ represent two oscillating 
dipoles with opposite directions, caused by the electron oscillation. 
(c) Near fields of the CSP (while far fields tend to zero along any direction).
\label{Fig1} }
\end{figure}

A metal has free electrons, which move/oscillate easily on the surface in 
response to external EM waves and thus may emit new wavelets.
But first note that a CSP corresponds to a surface-bound mode 
on the metal. 
If the oscillating charges emit light, how can the CSP be non-radiative? 
To clarify this ambiguity, let us see the Otto geometry in 
Fig.~\ref{Fig1}(a) as an example.\cite{CSP} 
At a specific incident 
angle $\theta_{\textrm{sp}}$ 
[greater than the critical angle $\arcsin(1/n_p)$ of the prism-vacuum 
interface], the incident wave can excite 
a CSP, which is a sinusoidal SCDW on the metal surface with a wavevector 
\begin{equation}
\label{CSP}
k_{\textrm{sp}} = K\left(1+1/\varepsilon_c\right)^{-1/2}, 
\end{equation}
where $\varepsilon_c$ is the 
permittivity of the metal and 
$K = 2\pi/\lambda$ ($\lambda$ the incident wavelength in vacuum). Here 
$\theta_{\textrm{sp}}$ must satisfy 
$n_p K \sin\theta_{\textrm{sp}} = \textrm{Re}(k_{\textrm{sp}})$, 
where $n_p$ is the refractive index of the prism. Under this condition, 
the incident energy is largely transferred to the CSP, giving rise to a 
reflection dip, as can be proved by Fresnel theory.\cite{CSP} 

Note that CSPs can be activated only on metals with 
Re$(\varepsilon_c)<0$ 
[and meanwhile Im$(\varepsilon_c)$ being small].\cite{CSP} 
The reason is that under this condition, the spatial period of the CSP 
satisfies 
\begin{equation}
\label{Sub-d}
d = 2\pi/\textrm{Re}(k_{\textrm{sp}}) < \lambda 
\end{equation}
based on Eq.~(\ref{CSP}). Therefore, the CSP is a 
\emph{subwavelength\/} charge pattern compared with the incident wavelength 
$\lambda$. Consider each period of the CSP in Fig.~\ref{Fig1}(a) as a scatter 
unit that emits new wavelets. Along any arbitrary direction $\theta$ $\neq 0$, 
the wavelets emitted from two adjacent units have a path difference 
\begin{equation}
\label{PathDiff}
\Delta S = d\sin \theta < d < \lambda, 
\end{equation}
i.e., the phase difference is less than $2\pi$. This means that the oblique 
wavelets can never be \emph{in phase}. Thus, they tend to cancel each other 
out in the far fields. The wavelets along the vertical direction $\theta = 0$, 
however, are in phase ($\Delta S = 0$), but viewed from a single period of the 
sinusoidal SCDW [Fig.~\ref{Fig1}(b)], each wavelet consists of two 
sub-wavelets with opposite electric fields ${\mathbf E}_1$ and 
${\mathbf E}_2$ that are also cancelled in the far fields. 
Therefore, all the emitted wavelets cannot escape the surface 
along any direction, so they form \emph{evanescent waves\/} near the 
surface [Fig.~\ref{Fig1}(c)]. This gives a simple picture why a CSP 
corresponds to a surface-bound mode. The CSP can thus 
propagate outside the prism-covered region in Fig.~\ref{Fig1}(a) 
without radiation loss, and the propagation distance solely 
depends on the absorption of the metal (Ohmic loss). 
Here it is obvious that media with 
$|\varepsilon_c| \rightarrow \infty$ (for perfect conductors) or 
Re$(\varepsilon_c) > 0$ do not support CSPs as the wavevector in 
Eq.~(\ref{CSP}) cannot satisfy Eq.~(\ref{Sub-d}) under these conditions. 

\begin{figure} 
\includegraphics[scale=1.0,angle=0]{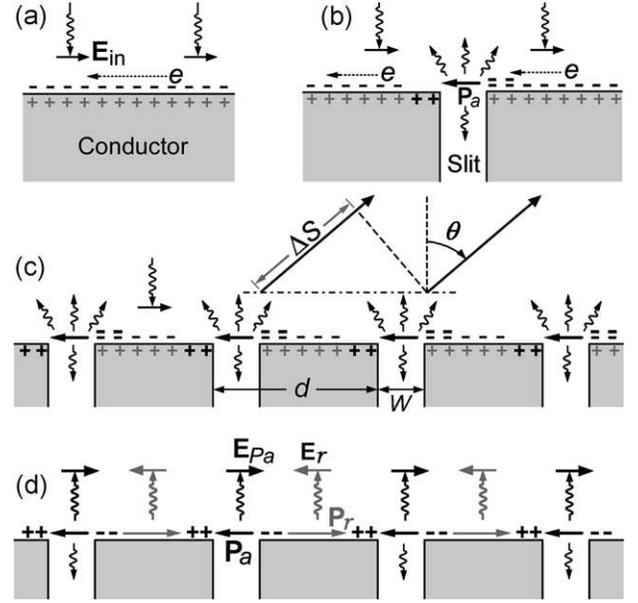}
\caption{Incident-wave-driven electron movement on (a) a flat conducting 
surface with no net charge, 
(b) a conducting surface with a single slit (net charges and their oscillation 
at the slit corners giving rise to a radiative light source), 
and (c) a conducting surface with a subwavelength slit array. 
(d) Periodic net charge pattern on the upper surface of (c). 
All the dipoles and electric fields have a common oscillating
factor $e^{i\omega t}$ ($t$ the time).
\label{Fig2} }
\end{figure}

Now we consider in Fig.~\ref{Fig2}(a) a plane wave incident on a conducting 
surface without the prism. 
For normal incidence, the incident electric field ${\mathbf E}_{\textrm{in}}$ 
drives free electrons on the surface to move homogenously. 
So there is no net charge, and the reflection obeys Fresnel 
equations.\cite{Born} 
In Fig.~\ref{Fig2}(b), a slit (or hole) is added. Apparently, the electron 
movement now can be impeded near the slit corner.
Here some electrons may move continuously to the vertical slit wall, but such 
movement corresponds to a $90^{\circ}$ deflection (acceleration) while the 
incident wave does not directly provide the necessary large driving force. 
So it is reasonable to assume that most of the moving electrons are stopped 
near one corner, while positive charges appear at the opposite corner because 
some electrons have moved away.
This leads to the formation of an electric dipole ${\mathbf P}_a$ at the 
slit opening. ${\mathbf P}_a$ oscillates with the incident wave with a time 
factor $e^{i\omega t}$ ($\omega$ the angular frequency of the incident wave), 
thus acting as a new \emph{light source\/} emitting wavelets. Such a process 
is in fact a Thomson scattering process in the optical frequency range.

Next we apply this process to the one-dimensional (1D) periodic slit array 
in Fig.~\ref{Fig2}(c). For simplicity, we assume 
that the grating is semi-infinite so that there is no feedback from below. 
Similar to Fig.~\ref{Fig2}(b), now each slit becomes a light source, 
but along any oblique direction $\theta \neq 0$, the wavelets emitted from 
two adjacent sources have a path difference $\Delta S = d\sin\theta$, 
where $d$ is the period of the slit array. 
For an incident wavelength $\lambda>d$,
Eq.~(\ref{PathDiff}) is satisfied again. Then the oblique wavelets are 
cancelled out in the far fields 
(destructive interference, similar to the absence of x-ray diffraction 
at non-Bragg angles), 
i.e., they also form evanescent waves near the surface. 
(This principle can also explain the fact that no light diffraction occurs 
from single crystal lattice, where the lattice constants are much smaller 
than the wavelength although the electrons 
still oscillate with the incident wave.) 

The charge pattern in Fig.~\ref{Fig2}(c) is  
similar to the CSP picture in Fig.~\ref{Fig1}, 
i.e., they are both \emph{subwavelength charge patterns\/} ($d < \lambda$).
However, there are two distinct differences. 
First, the 
CSP is a propagating wave with a specific wavevector 
determined by the metal's permittivity $\varepsilon_c$ in Eq.~(\ref{CSP}), 
while the charge pattern in Fig.~\ref{Fig2}(c) is a standing wave 
(but not sinusoidal) with the 
period \emph{always\/} equal to the grating period $d$. So the former is an 
intrinsic property of the metal (depending on $\varepsilon_c$) 
while the latter is a geometrical 
effect that can occur for any incident wavelength $\lambda>d$ 
and for any conducting materials containing free electrons 
[including perfect conductors and conductors with Re$(\varepsilon_c)>0$]. 
Second, as mentioned above, a CSP is a complete surface-bound mode. 
In contrast, the oscillating charge pattern in Fig.~\ref{Fig2}(c) is radiative 
along $\theta = 0$ ($\Delta S = 0$). This can be seen from Fig.~\ref{Fig2}(d), 
where we have discarded the oblique evanescent wavelets and added the 
dipoles ${\mathbf P}_r$ that are ignored in Fig.~\ref{Fig2}(c). In addition 
to the wavelet ${\mathbf E}_{Pa}$ emitted from ${\mathbf P}_a$, 
${\mathbf P}_r$ also emits a wavelet ${\mathbf E}_r$ along $\theta=0$ with a 
phase that is usually very close to that of the Fresnel reflected wave. 
So here we let ${\mathbf E}_r$ include Fresnel reflection 
for convenience in discussions. Then the wavelet emitted from a period 
consists of two sub-wavelets ${\mathbf E}_{Pa}$ and ${\mathbf E}_r$ 
along $\theta = 0$ with opposite directions (phases). 
But unlike Fig.~\ref{Fig1}(b), 
${\mathbf E}_{Pa}$ and ${\mathbf E}_r$ generally have different strengths, so 
they cannot completely offset each other. This leads to a propagating backward 
wave. Therefore, the charge pattern illustrated 
in Figs.~\ref{Fig2}(c) and \ref{Fig2}(d) is not a CSP, but one might call 
it a \emph{spoof SP\/} due to its similarities to the true CSP 
in Fig.~\ref{Fig1}(a).\cite{Pendry, Pendry2} 

From Fig.~\ref{Fig2} it is not difficult to obtain a general picture about 
light scattering from structured (or rough) conducting surfaces 
(either periodic or nonperiodic). When light is incident on a non-planar 
conducting surface, it drives the free electrons to move, but the movement 
can be impeded by the rough parts (particularly sharp edges) of the surface 
to form inhomogeneous oscillating charges, which become new light sources to 
emit wavelets. It is the interference between these wavelets that may 
give rise to anomalous reflection or scattering. 
In the following, we will numerically prove this mechanism in the simple and 
well studied case of periodic 1D gratings using the rigorous coupled-wave 
analysis (RCWA) technique.\cite{RCWA1, RCWA2} 

\section{RCWA of 1D lattice}
\label{SecIII}

For monochromatic waves in a nonmagnetic medium 
(permeability $\mu \equiv 1$), the electric and magnetic 
fields are coupled by Maxwell's equations (in c.g.s. units) 
\begin{eqnarray}
\label{MaxW1} 
\nabla \times {\mathbf E} &=& -iK{\mathbf H}, \\
\label{MaxW2}
\nabla \times {\mathbf H} &=& iK\varepsilon{\mathbf E},
\end{eqnarray}
where $K = 2\pi / \lambda$ and $\varepsilon$ is 
the \emph{effective permittivity}. 
The effective permittivity of a conductor can be expressed as 
$\varepsilon_c =  \varepsilon_c' - i4\pi \sigma/\omega$,
where $\varepsilon_c'$ is the regular permittivity and $\sigma$ is
the conductivity.\cite{Born} For perfect conductors,
$\sigma \rightarrow \infty$ so that Im$(\varepsilon_c) \rightarrow -\infty$ 
(which can also be derived from the Drude model of electrical conduction).
The divergence of Eq.~(\ref{MaxW2}) gives 
$\nabla \cdot (\varepsilon {\mathbf E}) = 0$, or
\begin{equation}
\label{ChargeDensity}
\nabla \cdot {\mathbf E} =
-[(\nabla \varepsilon) \cdot {\mathbf E}]/\varepsilon = 4\pi \rho,
\end{equation}
where $\rho({\mathbf r})$ is the \emph{bulk charge density\/}
(including both free and polarization-induced charges). 
In a modulated medium with varying $\varepsilon({\mathbf r})$, 
$\nabla \varepsilon \neq 0$, which generally leads to inhomogeneous charge 
densities $\rho({\mathbf r})$ according to Eq.~(\ref{ChargeDensity}).
Mathematically, $\varepsilon$ is discontinuous across a sharp interface 
(i.e., $\nabla \varepsilon \rightarrow \infty$), so
one has to use the \emph{surface charge density\/} 
$4\pi \tilde{\rho}_s = \delta E_z$ to describe the charge distribution on the 
interface, where $\delta E_z$ is the jump of the perpendicular electric field 
component across the interface.
As an exception, TE-polarization in a 1D structure 
satisfies $\rho \equiv 0$ since $(\nabla \varepsilon) \cdot {\mathbf E} = 0$, 
so here we ignore it.\cite{PRA}

\begin{figure} 
\includegraphics[scale=1.0,angle=0]{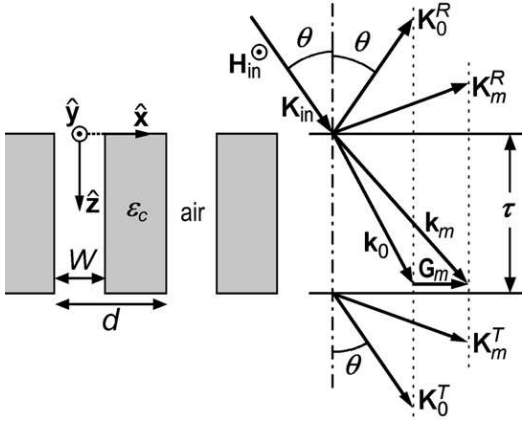}
\caption{Geometry of light diffraction from the 1D periodic grating. 
The two slit walls are located at $x = \pm W/2$ (plus any multiple of $d$). 
The vertical components
of the internal wavevectors ${\mathbf k}_m$ are generally complex vectors.
$\hat{\mathbf x}$, $\hat{\mathbf y}$, $\hat{\mathbf z}$ are unit vectors along 
the $x$, $y$, $z$ axes, respectively.
\label{Fig3} }
\end{figure}

From Eqs.~(\ref{MaxW1}) and ({\ref{MaxW2}) one can obtain
a second-order differential equation
$\nabla \times (\varepsilon^{-1}\nabla \times {\mathbf H}) = K^2 {\mathbf H}$, 
of which the Fourier transformation form for the 1D lattice in Fig.~\ref{Fig3}
is
\begin{equation}
\label{MaxW4}
K^2 {\mathbf H}_m = -\sum_n \zeta_{m-n} {\mathbf k}_m \times {\mathbf k}_n 
\times {\mathbf H}_n
\end{equation}
based on the Fourier expansions
$\varepsilon^{-1}(x) = \sum \zeta_m e^{-iG_m x}$
and 
${\mathbf H}({\mathbf r}) = 
\sum {\mathbf H}_m e^{-i{\mathbf k}_m \cdot {\mathbf r}}$,
where 
$G_m = 2\pi m/d$ ($m$ being integers),  
$\zeta_m = d^{-1}\int_0^d \varepsilon^{-1}(x) e^{iG_m x}dx$,
${\mathbf k}_m = {\mathbf k}_0 + G_m \hat{\mathbf x}$, and ${\mathbf k}_0$
is the forward wavevector.
Eq.~(\ref{MaxW4}) can be numerically solved by RCWA.
Here we briefly mention its main principles.

In Fig.~\ref{Fig3}, the incident wave is 
${\mathbf H}_{\textrm{in}} \exp(-i{\mathbf K}_{\textrm{in}} 
\cdot {\mathbf r})$ 
with 
${\mathbf K}_{\textrm{in}} = K(\sin\theta \hat{\mathbf x} 
+ \cos\theta \hat{\mathbf z})$. 
The forward wavevector ${\mathbf k}_0$ can be written as 
${\mathbf k}_0 = k_{0x}\hat{\mathbf x} + q \hat{\mathbf z}$ 
with $k_{0x} = K\sin\theta$, where $q$ is to be determined by 
the eigenequation. Then
the internal diffracted wavevectors have the form 
${\mathbf k}_m = k_{mx}\hat{\mathbf x}+q{\mathbf z}$
with $k_{mx} = k_{0x}+G_m$. 
Each diffraction order $m$ corresponds to two diffracted waves 
${\mathbf H}_m^R \exp(-i{\mathbf K}_m^R \cdot {\mathbf r})$ and 
${\mathbf H}_m^T \exp(-i{\mathbf K}_m^T \cdot {\mathbf r})$ 
above and below the grating, respectively.
Based on the conservation of the tangential wavevector components, we have 
\begin{eqnarray}
\label{WaveVectors}
K_{mx}^T \!\!&=&\!\! K_{mx}^R = k_{mx} = k_{0x} + G_m, \nonumber\\
K_{mz}^T \!\!&=&\!\! -K_{mz}^R = \left\{\!
\begin{array}{l}
(K^2 - k_{mx}^2)^{1/2}\ \textrm{ for } |k_{mx}| \leq K,\\
-i(k_{mx}^2 - K^2)^{1/2} \ \textrm{ otherwise}.
\end{array}
\right.
\end{eqnarray}
Here note that when $|k_{mx}|>K$, the corresponding external waves
become evanescent waves
$H_m^{R,T} e^{-ik_{mx}x}\exp[-(k_{mx}^2-K^2)^{1/2} |z|]$
along $\pm z$.
In particular, for normal incidence ($k_{0x} = 0$) and $\lambda>d$, 
all the external waves except for $m=0$ are evanescent,
$H_m^{R,T} e^{-2\pi i m x/d} \exp[-2\pi(m^2/d^2-1/\lambda^2)^{1/2}|z|]$. 
This is the mathematical description of the 
evanescent EM waves of the spoof SP described in Fig.~\ref{Fig2}.

For TM polarization, all the magnetic fields are parallel to 
$\hat{\mathbf y}$. If we retain $2M+1$ diffraction orders 
($0$, $\pm 1$, $\cdots$, $\pm M$),
Eq.~(\ref{MaxW4}) can be written as a $(2M+1)\times (2M+1)$ matrix 
eigenequation. 
From this eigenequation and the boundary conditions 
(continuity of the tangential electric and magnetic fields) 
at the two surfaces $z_s = 0,\tau$, one 
obtains $4M+2$ sets of eigenvalues $q_{\imath}$ and eigenmodes
$\{H_m^{\,\imath}\}$ ($m=-M$, $-M+1$, $\cdots$, $M$ and 
$\imath = 1$, $2$, $\cdots$, $4M+2$)
inside the grating and two sets of external fields $\{ H_m^{R} \}$
and $\{ H_m^{T} \}$
(see Refs.~\onlinecite{RCWA1} and \onlinecite{RCWA2} for details). 
Then the zero-order reflectivity and transmissivity are
$R_0 = |H_0^R/H_{\rm in}|^2$ and $T_0 = |H_0^T/H_{\rm in}|^2$, respectively. 
Meanwhile, the electric fields ${\mathbf E}_m^{\,\imath}$,  
${\mathbf E}_m^{R}$ and ${\mathbf E}_m^{T}$ 
(parallel to the $xz$ plane) are also obtained from (the Fourier 
transformation of) Eq.~(\ref{MaxW2}). Then the bulk charge density inside 
the grating ($0<z<\tau$) can be calculated from
\begin{equation}
\label{rhoV}
4\pi \rho({\mathbf r}) =-i \!\! \sum_{m=-M}^M \sum_{\imath = 1}^{4M+2}
({\mathbf k}_m^{\imath} \cdot {\mathbf E}_m^{\,\imath})
\exp ( -i{\mathbf k}_m^{\imath} \cdot {\mathbf r} ),
\end{equation}
where ${\mathbf k}_m^{\imath} = k_{mx}\hat{\mathbf x} 
+ q_{\imath} \hat{\mathbf z}$.
The surface charge density is
\begin{equation}
\label{rhoS1}
4\pi \tilde{\rho}_s (x) \! = \! - E_{{\rm in},z} 
e^{-ik_{0x}x} \!+\! \sum_m e^{-ik_{mx}x}
\!\biggl(\! -E_{mz}^R + 
\sum_{\imath} \! \! E_{mz}^{\,\imath}\! \! \biggr) 
\end{equation}
on the upper surface $z=0$ and
\begin{equation}
\label{rhoS2}
4\pi \tilde{\rho}_s (x) = \sum_m e^{-ik_{mx}x}
\biggl( E_{mz}^{T\tau}  
 - \sum_{\imath} E_{mz}^{\,\imath} \phi_{\imath} \biggr) 
\end{equation}
on the lower surface $z=\tau$, where $\phi_{\imath} = e^{-iq_{\imath} \tau}$
and $E_{mz}^{T\tau} = E_{mz}^T \exp(-iK_{mz}^T \tau)$.
[For large $\tau$, one may need to make the substitutions 
$H_m e^{-iq_{\imath} \tau} \rightarrow H_m^\tau$ and 
$H_m \rightarrow H_m^\tau e^{iq_{\imath} \tau}$
for Im$(q_{\imath})>0$
to avoid numerical overflow in computing
$e^{-iq_{\imath}\tau}$, where
$H_m^\tau$ is the corresponding internal wave amplitude at the lower surface.] 
For a semi-infinite grating ($\tau \rightarrow \infty$), only half of the 
eigenmodes with Im$(q_{\imath})<0$ are valid, so we only need to use the 
boundary 
conditions at the upper surface to compute the reflectivity and charge 
densities. Overall, RCWA is a first-principle method with the computation 
precision only depending on the number of diffraction orders ($2M+1$) 
retained, but note that it usually converges much slower in calculations 
of charges and near fields
than in calculations of (far-field) reflectivity and transmissivity. 

In calculating the bulk charge density $\rho(x, z)$ using 
Eq.~(\ref{rhoV}), we found that 
when a large number of diffraction orders are retained, $\rho(x, z)$ 
approaches a delta function across the walls, which means that ``bulk'' 
charges only exist on the slit walls, 
i.e., they are also surface charges.\cite{PRA}
Mathematically, we let $\rho(x\!=\!\pm W/2, \ z)$ represent the surface 
charge densities on the slit walls
(in arbitrary units). With sufficient orders retained, this approximation does 
not affect the shapes and phases of the real surface charge density curves on 
the slit walls.

\section{Semi-infinite gratings}
\label{SecIV}

In the above RCWA descriptions of light scattering from a 1D (or 2D) lattice,
the eigenmodes form pairs, each pair consisting of two eigenmodes with 
opposite (complex) vertical wavevectors $q$ and $-q$, respectively. 
As will be demonstrated later, one mode propagating along $-z$ corresponds 
to reflection from the bottom surface for finite $\tau$. This
mode can resonate with the forward-propagating one (along $+z$). 
In order to verify the picture in Figs.~\ref{Fig2}(c) and \ref{Fig2}(d) 
without the complication of the resonance,
we first consider a semi-infinite grating ($\tau \rightarrow \infty$) 
where the backward eigenmodes do not exist. 

\begin{figure} 
\includegraphics[scale=1.0,angle=0]{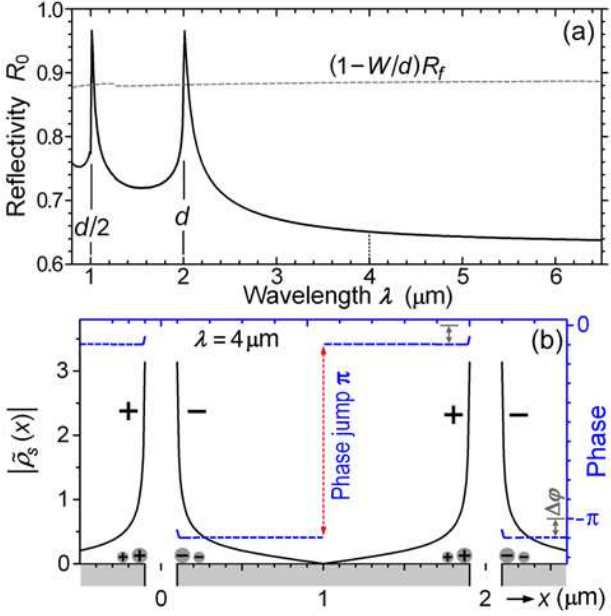}
\caption{(Color online) Reflectivity (a) and surface charge densities (b) of a semi-infinite 
gold grating with period $d = 2$ $\mu$m and 
slit width $W = 0.2$ $\mu$m. Normal incidence.
\label{Fig4} }
\end{figure}
 
Figure \ref{Fig4}(a) shows the reflectivity curve calculated with RCWA from a 
semi-infinite gold grating (practically $\tau>200$ $\mu$m) under normal 
incidence (with the frequency-dependent permittivity data of gold taken from 
Ref.~\onlinecite{HandBook}).
As a reference, the dashed line is the $(1-W/d)R_f$ curve with $R_f$ being 
the Fresnel reflectivity from a flat gold surface 
($R_f \simeq 0.98$ in the wavelength range $0.8 - 10$ $\mu$m) and $W$ the slit 
width. Compared with this reference curve, the anomalous reflection phenomenon 
from the grating is obvious. Generally the reflectivity $R_0$ is less than 
$(1-W/d)R_f$ except that near the Wood's anomalies $\lambda \simeq d/|m|$ 
($m\neq 0$ being integers), $R_0$ is close to unity.

Figure \ref{Fig4}(b) shows the charge density function $\tilde{\rho}_s (x)$ 
on the upper surface ($z=0$)
for an arbitrary wavelength in the $\lambda>d$ range. 
This function correctly shows that the incident wave indeed causes significant 
inhomogeneous charges on the grating surface with the charges highly 
accumulating near the slit corners, which is excellently consistent with the
charge distribution pattern predicted in Fig.~\ref{Fig2}(d).
At a time when ${\mathbf E}_{\textrm{in}}$ is toward $+x$ at $z=0$, we have 
predicted in Fig.~\ref{Fig2}(d) that the phase of the charge pattern is 
constant, equal to $-\pi$ (negative charges), on the left half surface 
$W/2 \leq x \leq d/2$, while for $d/2 \leq x \leq d-W/2$, the phase is $0$ 
(positive changes).
Figure \ref{Fig4}(b) shows that this prediction is largely correct except that 
the calculated phases are slightly displaced from the predicted phases $0$ 
and $\pi$ by $\Delta \varphi \simeq 0.1 \pi$ in most regions on the surface.
The phases near the slit corners are closer to the predicted values. 
Our calculations show that \emph{the charge patterns are nearly the same 
for any wavelength\/} $\lambda > 1.1 d$ with no resonance, and the phase 
shift $\Delta \varphi$ decreases with increasing $\lambda$, i.e., 
$\Delta \varphi \rightarrow 0$ for $\lambda \gg d$. Therefore,
the calculations indeed confirms the picture of charge accumulation and 
oscillation on the subwavelength lattice in Figs.~\ref{Fig2}(c) and 
\ref{Fig2}(d). Apparently, the period of the charge pattern in 
Fig.~\ref{Fig4}(b) is strictly equal to the lattice constant $d$ and is 
irrelevant to the dispersion property of CSPs in Eq.~(\ref{CSP}).
By performing RCWA calculations on gratings made of conductors with 
Re$(\varepsilon_c)>0$ or perfect conductors with
Im$(\varepsilon_c) \rightarrow -\infty$, we found that the main 
features of the charge patterns remain unchanged. 

To understand the anomalous reflection for $\lambda > d$ in 
Fig.~\ref{Fig4}(a), we may simply consider that,
as we mentioned before, the ${\mathbf E}_r$ wavelet in Fig.~\ref{Fig2}(d) 
consists of two contributions, 
${\mathbf E}_r = -(E_f + E_r')\hat{\mathbf x}$, 
with $E_f$ corresponding to regular Fresnel reflection 
($E_f = -\sqrt{R_f} E_{\textrm{in}}$)
and $E_r'$ corresponding to the emission of light from dipole 
${\mathbf P}_r$ along the backward direction $\theta = 0$.
Without charge accumulation, we have $E_{Pa} = E_r' = 0$ and the reflection 
should obey the Fresnel theory $R_0 = (1-W/d)R_f$.
When charges appear at the slit corners, it can be verified by RCWA that 
${\mathbf P}_a$ is strengthened much faster
than ${\mathbf P}_r$ (for $W<d/2$). Then the effective strength of 
${\mathbf E}_{Pa}$ becomes larger than $E_r'$. Consequently,
${\mathbf E}_{Pa}$ completely cancels $E_r'$, and also partially 
offsets $E_f$. Thus, the net effect is that
the overall reflectivity $R_0$ is smaller than $(1-W/d)R_f$. 

When $\lambda$ is reduced to be less than (or close to) the grating period 
$d$, some of the external wavevectors $K_{mz}^{R}$ 
in Eqs.~(\ref{WaveVectors}) become (or tend to be) real, and the corresponding 
diffracted waves become non-evanescent.
Then the diffraction effect appear, which can significantly changes the 
reflectivity (particularly at the Wood's anomalies $\lambda \simeq d/|m|$). 
The details in the diffraction range $\lambda < 1.1d$ are discussed in the 
Appendix since they no longer belong to subwavelength optics. 
But, it is worth emphasizing again here that the diffraction effect is absent 
for the entire long wavelength range $\lambda > 1.1d$, 
where the subwavelength charge patterns are always nearly the same as that in 
Fig.~\ref{Fig4}(b).

\section{Finite-thickness gratings}
\label{SecV}

In Fig.~\ref{Fig2} we have indicated that the dipole ${\mathbf P}_a$ 
also emits a wavelet in the slit toward $+z$ [see ${\mathbf E}_a$ in 
Fig.~\ref{Fig5}(a), which may also include a portion of
the incident wave directly transmitted into the slit]. 
Due to the waveguide constraint, ${\mathbf E}_a$ 
tends to be a plane wave inside the slit, i.e., 
${\mathbf E}_a \simeq E_a \exp(-ik_z z) \hat{\mathbf x}$,
where $k_z \simeq 2\pi/\lambda$. Similarly, this wave drives electrons 
\emph{on the slit walls\/} to oscillate, 
resulting in two SCDWs 
$\rho_a \exp(-ik_z z)$ and $-\rho_a \exp(-ik_z z)$ (with $\rho_a \propto E_a$) 
on the two opposite walls, respectively.
The SCDWs and the ${\mathbf E}_a$ wave propagate along $+z$ and attenuate 
gradually due to the absorption of the conductor. If the grating is extremely 
thick, these waves can be completely absorbed before reaching the bottom 
surface, which corresponds to the semi-infinite case. 

If the grating is thin enough, the SCDWs on the walls can reach the exit 
surface without significant absorption.
Then in a similar way, the moving charges can be impeded at the lower slit 
corners, leading to another large oscillating dipole ${\mathbf P}_b$, as shown 
in Fig.~\ref{Fig5}(a).\cite{PRA}
${\mathbf P}_b$ can give a strong feedback to the
upper surface by emitting a wavelet 
${\mathbf E}_b \simeq E_b \exp(ik_z z) \hat{\mathbf x}$
propagating upward. ${\mathbf E}_b$ also corresponds to two
SCDWs, $\pm \rho_b \exp(ik_z z)$, on the two walls, which are the back-bounced 
SCDWs of the $\pm \rho_a \exp(-i k_z z)$ waves by the bottom corners.
If ${\mathbf E}_b$ is \emph{in phase\/} with ${\mathbf E}_a$ at $z=0$, it
enhances ${\mathbf P}_a$. The enhanced ${\mathbf P}_a$ subsequently strengthens 
${\mathbf E}_{Pa}$, ${\mathbf E}_a$, ${\mathbf P}_b$, ${\mathbf E}_b$, and so 
on. Then a Fabry-Perot-like resonant state is
formed, with ${\mathbf E}_a$ and ${\mathbf E}_b$ forming a standing wave
${\mathbf E}_a + {\mathbf E}_b \propto \cos(k_z z) e^{i\omega t}$
in the slit. 
Under this condition, ${\mathbf E}_r$ is largely offset by ${\mathbf E}_{Pa}$ 
in the far fields, leading to minimized backward reflection. 
Figure \ref{Fig5}(b) shows the zero-order reflectivity curve of a gold grating 
with thickness $\tau = 4$ $\mu$m. Compared with Fig.~\ref{Fig4}(a), one can 
see that Figure \ref{Fig5}(b) indeed shows a number of reflection dips 
corresponding to Fabry-Perot resonance.

\begin{figure} 
\includegraphics[scale=1.0,angle=0]{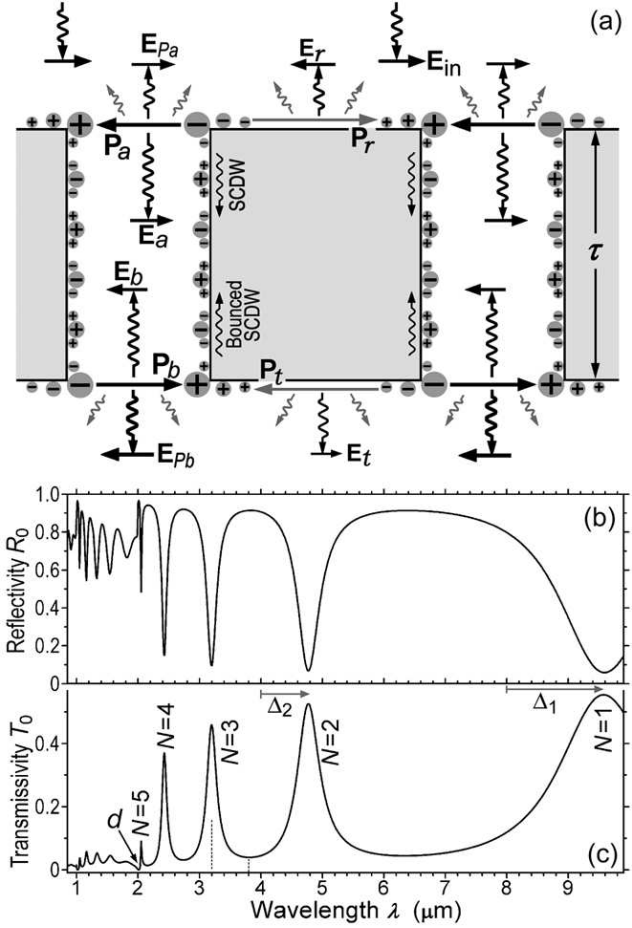}
\caption{(a) Process of charge oscillation-induced light emission, resonance 
and tranmssion through a conducting grating.
Reflectivity (b) and transmissivity (c) of a gold grating with parameters 
$d = 2$, $W = 0.2$, and $\tau = 4$ $\mu$m. 
\label{Fig5} }
\end{figure}

At the exit surface $z=\tau$ [Fig.~\ref{Fig5}(a)],
dipoles ${\mathbf P}_b$ and ${\mathbf P}_t$ also emit wavelets towards the 
outside of the slit. For $\lambda > d$, only the wavelets ${\mathbf E}_{Pb}$ 
and ${\mathbf E}_t$ can propagate along $+z$  
(while the oblique wavelets again form evanescent waves). Unlike the case 
above the upper surface where ${\mathbf E}_r$ contains
specular reflection, here wavelet ${\mathbf E}_t$ is purely emitted from 
dipole ${\mathbf P}_t$. For $W<d/2$, the strength of ${\mathbf P}_{b}$ 
(${\mathbf E}_{Pb}$) is much stronger than that of
${\mathbf P}_{t}$ (${\mathbf E}_t$), so the transmitted wave is dominated 
by ${\mathbf E}_{Pb}$. Consequently, the energy of the transmitted wave is 
highly localized near the exit opening. For long wavelengths
$\lambda \gg d$, such a ``near-field focusing'' effect can achieve a 
focusing width $W$ far smaller than $\lambda$, which has potential 
applications in nano-focusing/beaming, lithography, etc. In the far-field 
region, however, this effect
disappears as the transmitted beam becomes a plane wave.\cite{McNab}

At resonant wavelengths, since the strength of wavelet ${\mathbf E}_{Pb}$
is maximized, the zero-order transmissivity
$T_0$ is also maximized,\cite{PRA} as can be seen in Fig.~\ref{Fig5}(c), 
where each reflection dip exactly corresponds to a transmission peak
(also see similar results from finite-difference time-domain calculations 
in Ref.~\onlinecite{NGarcia}).

If the waves ${\mathbf E}_a$ and ${\mathbf E}_b$ are ideal 
plane waves with wavevector $k_z = 2\pi/\lambda$,
Fabry-Perot resonance should occur at $\lambda_N = 2\tau/N$ 
($N$ the resonance order) except that resonant peaks with 
$\lambda_N \leq d$ are suppressed by the diffraction effect.\cite{PRA} 
However, the actual resonance wavelength is always redshifted, 
$\lambda_N = 2\tau/N + \Delta_N$,
where the redshift $\Delta_N$ may vary (slowly) with $d$, $W$, $\tau$, 
and $\varepsilon_c$. 
One reason for the redshift is that the standing wave is distorted
near the two ends of the slit [see Fig.~\ref{Fig6}(b)]. 
One may refer to Ref.~\onlinecite{Suckling} 
for discussions of other possible mechanisms. 
Here note that due to the redshift, the spatial period of the 
(sinusoidal) SCDWs on the wall is less than the incident wavelength by 
approximately $\Delta_N$, i.e., they are also subwavelength charge
waves. Based on $\lambda_N \sim 2\tau/N$, thick (and highly conducting) 
gratings ($\tau \gg d$) have many resonance wavelengths in the 
non-diffraction range $\lambda >d$, as experimentally demonstrated
in Ref.~\onlinecite{rWent}. 

\begin{figure} 
\includegraphics[scale=1.0,angle=0]{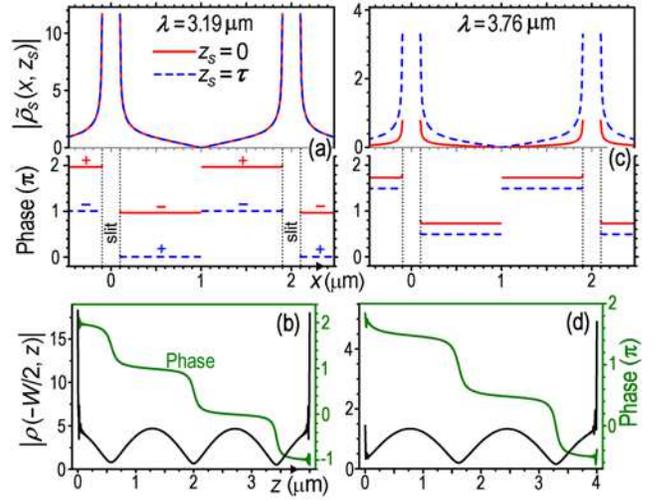}
\caption{(Color online) Charge patterns on the thin gold grating. 
(a) Charge densities on the two 
surfaces $z_s = 0$ and $\tau$ for the resonant wavelength 
$\lambda_3 = 3.19$ $\mu$m of 
peak $N=3$ in Fig.~\ref{Fig5}. (b) Nearly standing charge wave on the 
slit wall $x = -W/2$
at $\lambda_3$. Note that $\rho_v(W/2, z) \equiv -\rho_v(-W/2,z)$. 
(c) Surface charge densities at a non-resonant wavelength 
$\lambda = 3.76$ $\mu$m (corresponding
to the valley between peaks $N=2$ and $N=3$ in Fig.~\ref{Fig5}). 
(d) Charge wave on the slit wall $x=-W/2$ for
$\lambda = 3.76$ $\mu$m, where $z=0$ is no longer an anti-node 
of the standing wave.
\label{Fig6} }
\end{figure}

In Figs.~\ref{Fig6}(a) and \ref{Fig6}(b), the computed charge density 
distributions at resonant wavelength $\lambda_3$ 
are well consistent with the picture of Fig.~\ref{Fig5}(a). 
In Fig.~\ref{Fig6}(a), the two surface charge patterns 
$\left|\tilde{\rho}_s (x,\ z=0, \tau)\right|$
are very similar to that in Fig.~\ref{Fig4}(b), which confirms the existence 
of the large dipoles ${\mathbf P}_a$
and ${\mathbf P}_b$ in Fig.~\ref{Fig5}(a). Note that the charge densities in 
Figs.~\ref{Fig6}(a) and \ref{Fig4}(b) are in the same 
(arbitrary) unit. Therefore, the charge densities near the slit corners 
are much higher in Fig.~\ref{Fig6}(a) than in Fig.~\ref{Fig4}(b) due
to the Fabry-Perot resonance/enhancement. As also shown in Fig.~\ref{Fig6}(a), 
in thin gratings where
the attenuation of the charge density waves on the slit walls is negligible, 
the two SCDWs $\tilde{\rho}_s (x, \ z=0, \tau)$ are
almost identical except that for odd resonant orders $N$, they have a phase 
difference $\pi$. For relatively thicker gratings, the strength of 
$\tilde{\rho}_s (x, z = \tau)$ drops with increasing $\tau$. 
For $\tau > 200$ $\mu$m, $\tilde{\rho}_s (x, z = \tau)$ almost disappear while 
$\tilde{\rho}_s (x, z = 0)$ tends to be the same as that in Fig.~\ref{Fig4}(b).

Figure \ref{Fig6}(b) correctly reveals that on the slit walls, the charge 
density waves $\rho (x=\pm W/2, z)$ 
with approximately stepped phases are nearly standing waves.
Here the $\rho$ profile also shows high accumulation of charges at the slit 
corners $(x=\pm W/2, \ z=0,\tau)$ that
are (always) in phase with 
$\tilde{\rho}_s (x=\pm W/2, \ z=0, \tau)$.\cite{PRA}

In Fig.~\ref{Fig5}(a), if ${\mathbf E}_b$ not in phase with ${\mathbf E}_a$ 
(and ${\mathbf E}_{\textrm{in}}$) at $z=0$, 
it suppresses the strengths of ${\mathbf P}_a$ and ${\mathbf E}_{Pa}$ and 
influences their phases. Consequently the strengths of the charge waves on 
the slit walls are also reduced, leading to a weaker dipole ${\mathbf P}_b$ 
and weak transmissivity. This mechanism is clearly shown in
Figs.~\ref{Fig6}(c) and \ref{Fig6}(d) at a non-resonant wavelength.  
Compared with Figs.~\ref{Fig6}(a) and \ref{Fig6}(b), the charge densities 
at the slit corners $(x=\pm W/2, \ z=0, \tau)$ all drop significantly
for both $\rho$ and $\tilde{\rho}_s$, particularly at the upper corners. 
Meanwhile, the phases of the charge waves are also altered so that no 
resonance is formed. As stated above, without surface charges, 
the reflectivity from the upper surface should be the Fresnel 
reflectivity $R_0 = (1-W/d)R_f$. Here one can see
from Figs.~\ref{Fig6}(c) and \ref{Fig6}(d)
that at non-resonant wavelengths, the strengths of ${\mathbf P}_a$ 
and ${\mathbf P}_r$ at the upper surface 
are very small, and then the
reflectivity $R_0$ in Fig.~\ref{Fig5}(b) is indeed very close to 
$(1-W/d)R_f \simeq 0.9$ 
in most of the non-resonant wavelength range.
For the same wavelength, the non-resonant reflectivity in Fig.~\ref{Fig5}(b) 
is much stronger than that in Fig.~\ref{Fig4}(a) where charge oscillation is 
heavily involved. This further proves the essential role charge oscillation 
plays in extraordinary light scattering from metamaterials.

As mentioned above, perfect conductors with 
Im$(\varepsilon_c) \rightarrow -\infty$ and conductors 
with Re$(\varepsilon_c) >0$ 
do not support CPSs. However, we have demonstrated in 
Ref.~\onlinecite{PRA} that 1D gratings with Re$(\varepsilon_c) >0$ may still 
show similar extraordinary light transmission although the transmissivity 
is relatively lower.\cite{Lezec1} Here we use RCWA to simulate 
the transmission through a nearly perfectly conducting
grating with a large constant imaginary permittivity 
$\varepsilon_c \equiv - i100\ 000$. 
Based on this value, the wavevector $k_{\textrm{sp}}$
in Eq.~(\ref{CSP}) is almost accurately equal to $K$, so CSPs should not 
exist. However, our calculations show that all the major  
properties in this case are almost identical to that of regular metallic 
gratings. For example, Figure \ref{Fig7}(a) shows the transmissivity curve 
calculated with the same geometrical
parameters in Fig.~\ref{Fig5}(c), while Figure \ref{Fig7}(b) shows the 
charge densities on the grating surfaces for the third-order resonance peak. 
Compared with the reflectivity curve in 
Fig.~\ref{Fig5}(c) and the charge density distribution in
Fig.~\ref{Fig6}(a), Figure \ref{Fig7} apparently indicates that the light 
scattering mechanisms for the perfect-conductor case are the same, and thus 
are irrelevant to CSPs. In fact, the resonant transmissivity peaks and the 
charge densities in Fig.~\ref{Fig7} are averagely higher than those for the 
gold grating, indicating that high conductivity can (significantly) enhance 
the extraordinary scattering effects. 

\begin{figure} 
\includegraphics[scale=1.0,angle=0]{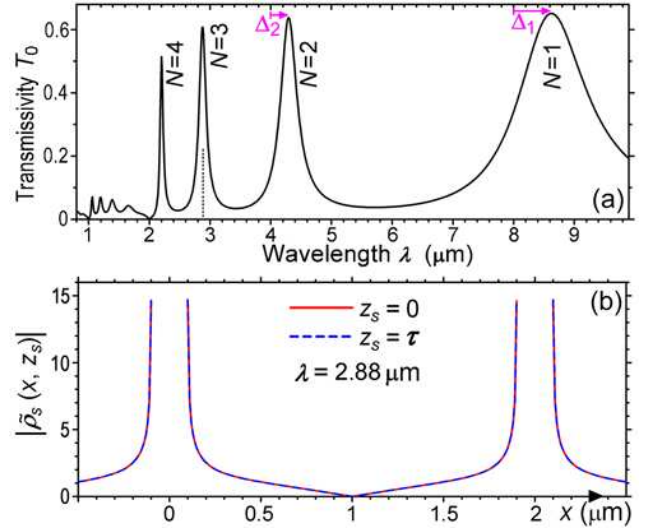}
\caption{(Color online) Light scattering from a nearly perfectly conducting grating. 
$\varepsilon_c \equiv -i100\ 000$ (compared with $\varepsilon_c = -2834-i1333$ 
for gold at $\lambda = 9.9$ $\mu$m).
$d = 2$, $W = 0.2$, and $\tau = 4$ $\mu$m. Normal incidence. 
(a) Transmissivity curve. 
Compared with Fig.~\ref{Fig5}(c), here the redshifts ($\Delta_N$) of 
the resonant wavelengths are smaller
(so peak $N=5$ below $\lambda = d$ is truncated)
and the 
resonant transmissivities are larger.
\label{Fig7} }
\end{figure}

As demonstrated in refs.~\onlinecite{PRA} and \onlinecite{APL}, extraordinary 
transmission or scattering through 2D hole arrays involve the same 
mechanisms of light emission and interference except 
that the tunneling of the SCDWs through the holes is different.
The details of oblique incidence geometry will be presented elsewhere, but the 
charge oscillation principle is similar. 

\section{Nonperiodic structures}
\label{SecVI}

From the above demonstrations, it becomes obvious that charge 
oscillation-induced light emission and interference 
are a fundamental and \emph{universal\/} mechanism underlying 
various extraordinary light scattering processes 
from conducting structures although these processes may also 
involve other mechanisms simultaneously (e.g. cavity resonance). 
The only basic requirement for 
this mechamism to work is that the structure 
have free electrons. So this mechanism applies 
to structures of  metals, perfect conductors, conductors with
Re$(\varepsilon_c)>0$, semiconductors, etc, but high conductivity 
can significantly enhance the anomalous scattering effects. 

This mechanism also applies to nonperiodic structures.\cite{NonPeriodic}
From Fig.~\ref{Fig2}(b) one can see that an isolated
single slit also acts as a light source. If the conducting plate has a finite 
thickness, the exit opening of the slit at the lower surface becomes another 
strong light source at a Fabry-Perot resonant wavelength,
emitting a transmitted beam below the plate.\cite{Suckling} 
Compared with the periodic slit array in Fig.~\ref{Fig5}(a), 
the waves emitted from a single slit have no interference and thus are 
completely divergent.
If the slit is surrounded by periodic grooves on the entrance surface,
as shown in Fig.~\ref{Fig8}(a), each groove now acts as a light source. 
Under the conditions that the groove period is less than the incident 
wavelength and that the Fabry-Perot-like resonance can be achieved 
simultaneously in both the grooves and the slit, the EM fields above the 
upper surface become similar to those in Fig.~\ref{Fig5}(a) with the
oblique waves forming evanescent modes. 
Most importantly, the enhanced backward wavelets from 
the light sources [${\mathbf E}_{Pa}$ in Fig.~\ref{Fig5}(a)] 
significantly reduce the Fresnel reflection (${\mathbf E}_r$). 
Accordingly, the backward reflection is reduced while the transmission through 
the slit can be greatly enhanced (by up to 2 orders 
in Ref.~\onlinecite{SlitGroove}). 
But, the transmitted beam below the plate is still divergent.

Now if similar grooves are made on the exit surface, the wavelets 
emitted from the exit opening of the slit also
drives free electrons to form oscillating dipoles at the openings.
Thus, they also become light sources. Similarly, the oblique 
wavelets emitted from the slit opening and the grooves on
the exit surface tend to form evanescent waves, giving rise to a narrow 
and directed transmitted beam below the slit opening.
However, the dipoles on the upper entrance surface are formed and 
driven mainly by the 
wide incident wave while those on the exit
surface result only from the wavelets emitted from the dipole of the 
single slit. Therefore, the strengths of the light sources on the exit surface 
decrease quickly with increasing distances of the grooves from the slit.
Due to this reason, the transmitted beam cannot be completely 
collimated since the oblique wavelets cannot be completely
suppressed. Meanwhile, the grooves on the exit surface
have little influence on the overall transmission efficiency.\cite{SlitGroove} 

Obviously, the picture illustrated in Fig.~\ref{Fig8}(a) 
can also explain enhanced light transmission and directed
nanobeaming through a single aperture surrounded by circular 
grooves in the 2D case except that the cavity resonance mechanisms 
in the aperture and in the circular grooves may be
different and the directions and distributions of the light sources near the 
groove edges are more complicated.\cite{r3} 
 
As another nonperoidic structure example, it is known that when one end of a 
conducting nanowire is illuminated by a narrow-wavefront beam (with 
the electric field ${\mathbf E}_{\textrm{in}}$ being polarized along the 
wire), the other end that is not illuminated can emit light, 
as schematically illustrated in Fig.~\ref{Fig8}(b). 
The common explanation of this phenomenon is that light is transferred by CSPs 
on the wire surface.\cite{Nanowire1} However, it is found that this phenomenon 
is more pronounced in the low-frequency (e.g. terahertz) range, where most 
metals become nearly CSP-free perfect conductors.
In particular, Wang and Mittleman\cite{Nanowire2} have experimentally 
demonstrated that in the terahertz range ($\lambda \sim 1$ mm), 
the wave modes on metallic nanowires have a dispersion trend that is  
opposite to that of CSPs. 

In fact, according to our charge oscillation picture, the basic mechanism 
underlying light transfer on nanowires is very simple. In Fig.~\ref{Fig8}(b), 
the incident wave drives free electrons
near the left input end to oscillate. The agitated electrons then 
propagate outside the illumination area towards the right side as a SCDW. 
At the other end the propagating charge wave is discontinued, 
giving rise to a strong charge accumulation there.
The oscillation of these charges them emit new light near the 
exit end. Meanwhile, the charge wave is bounced back. When
the bounced charge wave is in phase with the forward wave 
(and the incident wave) at the input end, Fabry-Perot
resonance occurs. This is very similar to the
charge movement on the slit walls in Fig.~\ref{Fig5}(a). 
In general, the resonant wavelength here also has a redshift. 
Accordingly, the standing charge wave on the wire surface is 
a subwavelength wave ($d<\lambda$), and based on 
Fig.~\ref{Fig1}, it generates little
radiation loss when propagating on the wire. The Fabry-Perot 
resonance and the subwavelength charge patterns 
(proportional to the strengths of the near fields) indeed have 
been demonstrated both experimentally
and theoretically.\cite{Nanowire1, Nanowire2, Nanowire3}.  
Interestingly, our calculations show that a charge wave 
propagating in the unilluminated region of a flat/straight 
conducting surface (including the slit wall and the straight wire)
is always a subwavelength SCDW, which indicates that the general 
SP picture elaborated in the literature, light $\rightarrow$ subwavelength 
SCDWs $\rightarrow$ light, is indeed correct except that the SCDWs are 
spoof SPs and do not necessarily have
the dispersion property of Eq.~(\ref{CSP}). 

\begin{figure} 
\includegraphics[scale=1.0,angle=0]{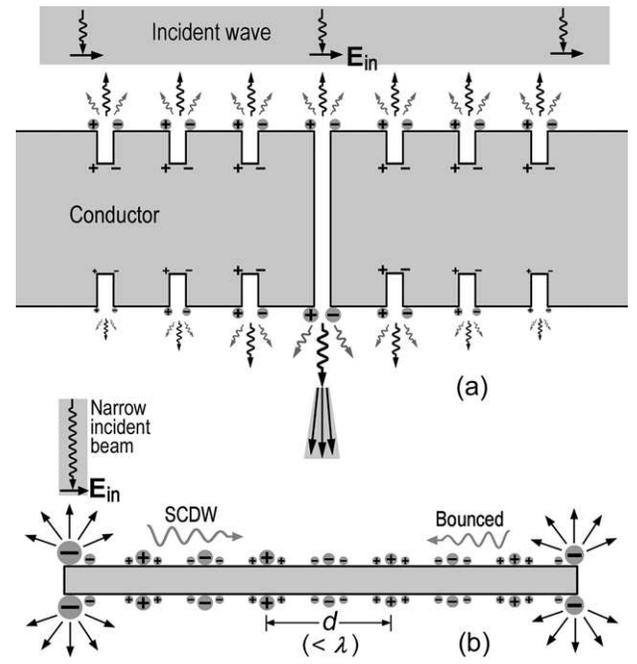}
\caption{(a) Schematic representation of enhanced light transmission 
and directed 
nanobeaming through a single slit surrounded by grooves. 
(b) Transfer of light on a conducting nanowire by
subwavelength SCDWs. 
\label{Fig8} }
\end{figure}

According to this picture, the conductivity of the nanowire
is the dominant factor determining the propagating distance of the 
charge waves and the efficiency of light transfer.
This explains the remarkably high transfer efficiency in the 
long-wavelength range where most metals are highly conducting.
To further confine the 
near fields so as to reduce the radiation loss 
(caused by possible deviations
of the actual charge waves from ideal subwavelength 
standing waves), one may activate charge waves in
grooves and guide them to propagate inside the grooves. 
In these cases, the charge waves are channel spoof 
SPs \cite{ChannelSP} that may have longer propagating distances.

From Fig.~\ref{Fig8}(b) it is obvious that to achieve high transfer 
efficiency, the diameter (vertical dimension) 
of the wire should
be much smaller than the incident wavelength so that the agitated 
charge waves on the top and bottom of the
wire have nearly the same phase. Otherwise, the charge waves with
different phases will quickly mix together and thus offset each 
other outside the illuminated region, 
leading to a short propagation distance. This explains why light transfer is 
remarkable on \emph{nanowires}. 
With respect to this effect, it is expected that a thin conducting slab would 
be more efficient since it can enhance the input coupling efficiency and 
reduce the electrical resistance without causing phase differences.

Note that in Fig.~\ref{Fig2}(c), when the moving electrons are stopped at the 
slit corners, they also have a tendency to be bounced back, similar to the
moving electrons on the nanowires in Fig.~\ref{Fig8}(b) [and on the slit walls 
in Fig.~\ref{Fig5}(a)]. 
The difference in Fig.~\ref{Fig2}(c) is that the bounced back charges
are suppressed by the incident electric field ${\mathbf E}_{\textrm{in}}$ 
since the driving force provided by
${\mathbf E}_{\textrm{in}}$ is always opposite to this tendency, 
while on the nanowire of Fig.~\ref{Fig8}(b), ${\mathbf E}_{\textrm{in}}$ 
is absent except at the input end. 

\section{Summary}
\label{SecVII}

By numerically calculating the SCDWs on gratings, we have 
demonstrated that the incident wave can drive free electrons to
accumulate and oscillate near the slit corners to form new light sources.
These light sources then emit new wavelets. 
For periodic subwavelength structures ($d<\lambda$), the
oscillating charges form subwavelength charge patterns (i.e., spoof SPs) and 
the wavelets emitted from them destructively interference with each 
other to form evanescent wave modes near the  
surfaces. Usually combined with other mechanisms (e.g. Fabry-Perot or cavity 
resonance), the spoof SPs can lead to anomalous light reflection, 
transmission, or scattering. The spoof SPs are mainly a geometrical effect 
and generally do not have the dispersion properties of CSPs. 
Note that in the literature, the SP-like modes on metamaterials with finite 
conductivity were widely assumed to be CSPs,
while only those on perfectly conducting structures were 
believed to be spoof SPs. Here we have demonstrated 
that they are all spoof SPs. 
(For transmission of acoustic waves through gratings,\cite{Ming,Ming2} the 
counterpart of charge oscillation
is the mechanical vibration of the structured medium, particularly near 
the corners and edges, that emits acoustic wavelets.)

We also illustrated that the same mechanism of charge 
oscillation-induced light emission and interference applies to 
\emph{all structures with free electrons\/}
(including perfect conductors and nonperiodic structures). 
Thus, the spoof SP picture 
represents a basic and universal mechanism of light scattering from 
conducting nanostructures.
The guideline provided by this mechanism is that
in designing novel nano-metamaterial devices, there is no CSP excitation 
constraint, but one needs to precisely design the geometrical parameters 
of the devices so as to accurately control the locations of 
the new light sources (including maximizing the strength of 
various resonance processes if involved) and their interference. 
Meanwhile, choosing highly conducting materials is generally 
another requirement for enhancing the
anomalous scattering effects.

\section*{ACKNOWLEDGEMENTS}

This work was supported by grants
from the NSFC (Grant Nos. 10625417, 50672035, and
10874068), the MOST of China (Grant Nos. 2004CB619005
and 2006CB921804), and partly by the ME of China and
also Jiangsu Province (Grant Nos. NCET-05-0440 and
BK2008012).
X.R.H was supported by the U.S. Department of Energy, Office 
of Science, Office of Basic Energy Sciences, 
under Contract No. DE-AC-02-98CH10886. We thank Yong Q. Cai for helpful
discussions.

\appendix*
\section{Charge patterns in the diffraction range for 1D gratings}
\label{Append}

One might intuitively think that for normal incidence with the electric fields 
${\mathbf E}_{\textrm{in}}$ parallel to the surface in Fig.~\ref{Fig2}(c),
the charge pattern on the surface should remain the same even 
if $\lambda \leq d$, 
as ${\mathbf E}_{\textrm{in}}$ provides the same driving force.
However, the free electrons on the grating surface are driven by both the 
incident wave and the wavelets emitted from the charges/dipoles. The latter 
may significantly influence the charge distribution (to achieve 
self-consistency of the system) when $\lambda \leq d$.

Figure \ref{Fig9} shows the charge patterns in the short-wavelength range 
($\lambda$ close to or below $d$)
for the semi-infinite grating of Fig.~\ref{Fig4}(a).
To understand these patterns, recall in Sec.~\ref{SecIII} that  
the $\pm m$th-order wave components ($m>0$) above the grating 
can be written as ${\mathbf E}_m^R e^{- 2\pi i mx/d} \exp(-iK_{mz}^Rz)$ and
${\mathbf E}_{-m}^R e^{ 2\pi i mx/d} \exp(-iK_{mz}^Rz)$, respectively for 
normal incidence, where
$K_{mz}^R = - 2\pi (1/\lambda^2 - m^2/d^2)^{1/2}$.
By symmetry, $|{\mathbf E}_m^R| = |{\mathbf E}_{-m}^R|$. 
So these two waves form a standing wave 
$\bm{\mathcal E}_m \sin(2\pi mx/d) \exp(-iK_{mz}^Rz)$. Accordingly, there
exists a standing SCDW $\tilde{\rho}_{s,m} \sin(2\pi mx/d)$ corresponding to 
this wave on the metal surface (but discontinued in the slit gap).
For $\lambda>d/m$, the propagating factor $\exp(-iK_{mz}^Rz)$ of the 
$\bm{\mathcal E}_m$ wave becomes a decaying factor $\exp(-\beta_m|z|)$ along 
$-z$, where $\beta_m = 2\pi(m^2/d^2-1/\lambda^2)^{1/2}$ is the
decaying coefficient. Therefore, the $\bm{\mathcal E}_m$
wave is an standing evanescent wave under this condition. 
For $\lambda >d$ (so that $\lambda > d/m$),
all the $\bm{\mathcal E}_m$ waves except for $m=0$ are evanescent, so they
all have weak strengths. The charge pattern in Fig.~\ref{Fig4}(b) is 
the collective contributions from a large number of charge wave components
$\tilde{\rho}_{s,m} \sin(2\pi mx/d)$, and this pattern (as well as the 
collective near fields of the wavelets emitted from
the dipoles) coincides with the grating lattice.

\begin{figure} 
\includegraphics[scale=1.0,angle=0]{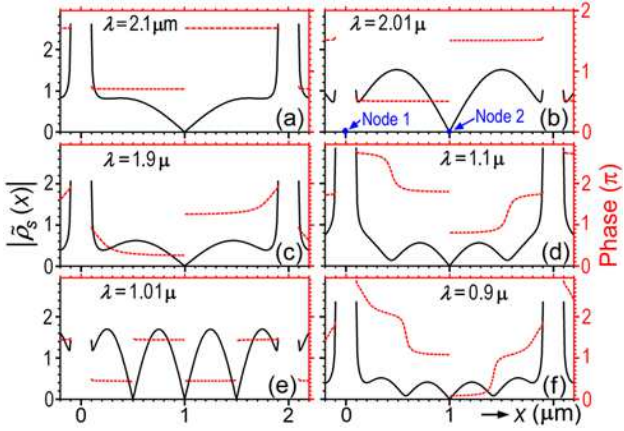}
\caption{(Color online) Diffraction affected charge densities 
on the surface $z = 0$ of a semi-infinite gold grating.
$d = 2$ and $W=0.2$ $\mu$m. Dashed lines are the phases. 
Normal incidence.
\label{Fig9} }
\end{figure}

When $\lambda$ decreases toward $d$ from above, the first-order 
$\bm{\mathcal E}_1$ wave tends to be strengthened as its decaying coefficient 
$\beta_1$ decreases. Compared with Fig.~\ref{Fig4}(b), one can see in 
Fig.~\ref{Fig9}(a) the appearance of the standing SCDW component 
$\tilde{\rho}_{s,1} \sin(2\pi x/d)$ superimposed on the overall charge pattern 
that is similar to that in Fig.~\ref{Fig4}(b).
(Here the charges are influenced by the near fields even if the 
$\bm{\mathcal E}_1$ wave is evanescent in the far fields.) 
Meanwhile, the phase shift $\Delta \varphi$ indicated in Fig.~\ref{Fig4}(b) 
increases to $-0.28\pi$ in Fig.~\ref{Fig9}(a) due to the phase of   
the complex constant $\tilde{\rho}_{s,1}$.  

The wavelength $\lambda = d$ corresponds to the first-order 
Wood's anomaly.\cite{PRA} Under this condition,
the wavelets emitted from two adjacent units of the grating along the 
horizontal direction $\theta = \pi/2$ have a phase difference of 
$2\pi$ according to Fig.~\ref{Fig2}(c), which corresponds to a special 
\emph{Bragg diffraction condition}. 
Then the strength of $\bm{\mathcal E}_1$ wave is maximized and becomes 
outstanding among other components, as can be verified by RCWA. 
Therefore, the charge pattern in Fig.~\ref{Fig9}(b) is dominated by the 
corresponding charge wave $\tilde{\rho}_{s,1} \sin(2\pi x/d)$. 
Here one (virtual) node (Node 1) of this standing SCDW is located at 
($x = 0,\ z = 0$), the middle of the slit [while the other node (Node 2) 
is at ($x=d/2, \ z=0$)]. Node 1 significantly suppresses the charge densities 
at the two slit corners (because the nodes of a standing wave have zero 
amplitudes), so the strength of the dipole ${\mathbf P}_a$ 
[see Figs.~\ref{Fig2}(c) and \ref{Fig2}(d)] is minimized. As discussed in 
Figs.~\ref{Fig6}(c) and \ref{Fig6}(d), when the strength of ${\mathbf P}_a$ 
is very small, the reflectivity should approach the Fresnel reflectivity 
$(1-W/d)R_f$. But note that in Fig.~\ref{Fig9}(b), the charge pattern 
$\tilde{\rho}_{s,1} \sin(2\pi x/d)$ also contributes \emph{positively\/} to 
the reflectivity [equivalent to the increase of $E_r'$ in Fig.~\ref{Fig2}(d) 
although it has a phase difference $\simeq \pi/2$ from the incident wave]. 
This is the reason why in Fig.~\ref{Fig4}(a) the reflectivity
at Wood's anomaly wavelengths is even higher than $(1-W/d)R_f$.
Note that the first-order Wood's anomaly
in Fig.~\ref{Fig4}(a) is slightly red-shifted from $d = 2$ to $2.01$ $\mu$m.

When $\lambda$ decreases to $1.9$ $\mu$m in Fig.~\ref{Fig9}(c), 
the $\bm{\mathcal E}_1$ wave is still a propagating mode, but its strength 
decreases as the diffraction condition deviates from the Bragg condition. 
Meanwhile, the $\tilde{\rho}_{s,2} \sin(4\pi x/d)$ wave begins to gain 
strength, which modifies the stepped phase profile. When $\lambda$
further decreases toward $d/2$ in Fig.~\ref{Fig9}(d), 
the $\tilde{\rho}_{s,2}\sin(4\pi x/d)$ wave becomes appreciable. 
At the second-order Wood's anomaly $\lambda =1.01$ $\mu$m (also slightly 
red-shifted from $d/2$), this wave is maximized [Fig.~\ref{Fig9}(e)]. For 
$\lambda < d/2$ [Fig.~\ref{Fig9}(f)], the third-order wave 
$\tilde{\rho}_{s,3} \sin(6\pi x/d)$ begins to show strengths, and so on. Thin 
gratings have similar properties, but these effects are meanwhile mixed with 
the resonance in the slits.

Overall, we may call the short-wavelength range $\lambda < 1.1 d$ 
the \emph{diffraction range}, where the diffraction effect of the 1D lattice 
appear, particularly at the Wood's anomalies. In the diffraction process, it 
is interesting that the incident energy is largely back reflected rather than
diffracted although the corresponding diffracted waves become non-evanescent, 
as can be seen from Fig.~\ref{Fig4}(a). This is also true in 
Figs.~\ref{Fig5}(b) and \ref{Fig5}(c) for the thin grating, where the 
reflectivity is averagely very high in the entire diffraction range while the 
transmissivity (particularly at the Fabry-Perot resonant transmission peaks) 
is much lower than that in the non-diffraction range $\lambda > 1.1d$.

\end{document}